\documentclass[12pt,preprint]{aastex}

\shorttitle{The Arecibo Intermediate Galactic Latitude Survey}
\shortauthors{Navarro, Anderson and Freire}

\begin{document}

\title{The Arecibo 430-MHz Intermediate Galactic Latitude Survey:
Discovery of Nine Radio Pulsars}

\author{J. Navarro,\altaffilmark{1}
  S. B. Anderson ,\altaffilmark{2} \and
  P. C. Freire \altaffilmark{3}}
\medskip
\affil{\altaffilmark{1}Schlumberger, Cambridge Research Ltd, High
Cross, Madingley Road, Cambridge, CB3 0EL, UK}
\affil{\altaffilmark{2}Caltech 18-34, Pasadena
CA 91125-3400, USA}
\affil{\altaffilmark{3}NAIC, Arecibo Observatory, HC 03 Box 53995,
Arecibo PR~00612, USA}

\bigskip

\begin{abstract}
We have used the Arecibo Radio Telescope to search for millisecond
pulsars in two intermediate Galactic latitude regions ($7^\circ < | b
| < 20^\circ$) accessible to this telescope. For these latitudes the useful
millisecond pulsar search volume achieved by Arecibo's 430-MHz beam is
predicted to be maximal. Searching a total of 130 square degrees,
we have discovered nine new pulsars and
detected four previously known objects. We compare the results of this
survey with those of other 430-MHz surveys carried out at Arecibo and
of an intermediate latitude survey made at Parkes that included part
of our search area; the latter independently found two of the nine
pulsars we have discovered.

At least six of our discoveries are isolated pulsars with ages
between 5 and 300 Myr; one of these, PSR~J1819+1305,
exhibits very marked and periodic nulling. We have also found a
recycled pulsar, PSR~J2016+1948. With a rotational period of 65
ms, this is a member of a binary system with a
635-day orbital period. We discuss some of the the properties of this
system in detail, and indicate its potential to provide a test of the Strong
Equivalence Principle.
This pulsar and PSR~J0407+16, a similar system now being
timed at Arecibo, are by far the best systems known for such a test.
\end{abstract}


\section{Introduction}\label{sec:intro}

Most of the pulsars found with the Arecibo radio telescope have been
discovered in blind surveys. Since 1991, a series of 430-MHz surveys,
which take advantage of the unparalleled gain of the 430-MHz line feed
(19 K/Jy) have been carried out. The papers containing the results of
most of these have already been published
(Nice, Fruchter, \& Taylor 1995, Camilo, Nice, \& Taylor 1996, Camilo
{et al.}  1996, Foster {et al.}  1995, Ray {et al.}  1996, {Lommen}
{et al.}  2000). Those surveys, together with processing of data taken
post-1997 ({Lorimer} {et al.}  2002, McLaughlin {et al.} 2003,
Chandler 2002), have uncovered a total of 113 pulsars. Of these,
 19 are recycled.

The optimal region for these surveys has always been considered to be
the Galactic plane, despite the high sky temperatures involved.
This was specifically the target region of the earliest surveys
(Hulse \& Taylor 1974, Hulse \& Taylor 1975a, Hulse \& Taylor 1975b).
The pulsars normally found in this region of sky are slow rotators
with relatively large Dispersion Measures (DM). They present a large
dispersive smearing across each of the channels of the back-ends used
to carry out the surveys. Such smearing makes the detection of
millisecond pulsars (MSPs) with rotational periods less than a few
times the timescale of the smearing a difficult proposition.

Dispersive smearing could in principle be
eliminated by using narrower channels, but at some point a more
fundamental problem is found, that of interstellar scattering. This is
the broadening of any pulsed signal due to multi-path propagation, and
can only be eliminated by observing at higher frequencies, where the
pulsars are intrinsically fainter and the telescope beam smaller.

\begin{figure}
\plotone{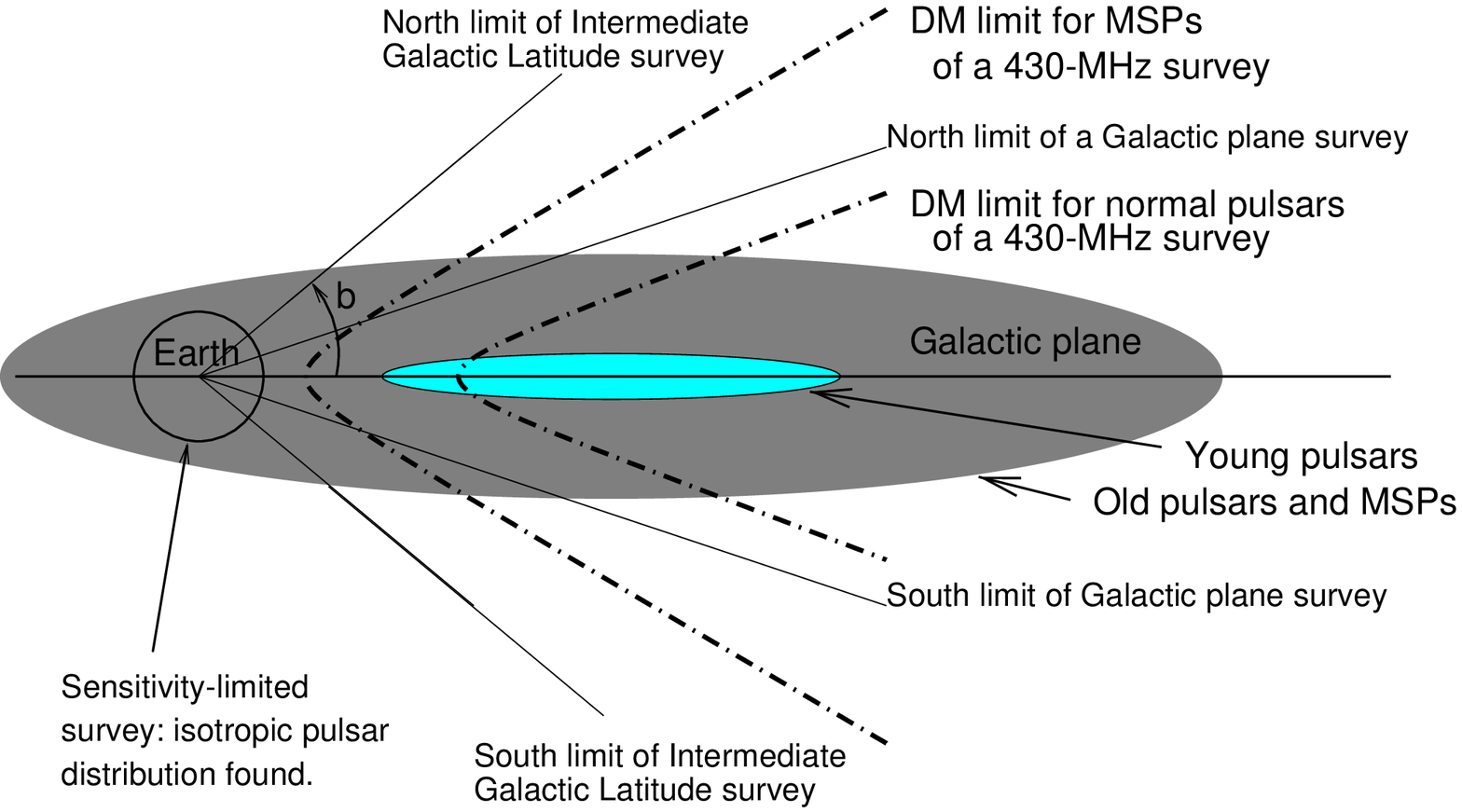}
\caption{Schematic cross section of the Galaxy at the
Galactic longitudes being probed, with some vertical exaggeration for
clarity. Both the ILS and the Galactic surveys
 are probing essentially the same
``normal'' pulsar population; mainly old ``normal'' pulsars, with ages
of tens to hundreds of Myr. The main difference between the two should
be the higher fraction of recycled objects for the ILS survey. The telescope
can see beyond the scale height of the pulsar population. The
consequence of this is that the pulsars detected in ILS are, on average,
less numerous and closer to the Earth than those found in the Galactic
surveys. Because of the large concentrations of plasma
near the Galactic plane, a survey at 430 MHz cannot detect young
pulsars, as they lie beyond its DM limit. The Parkes Multi-Beam survey
was designed to probe this region at 1400 MHz, discovering many young
pulsars. \label{fig:population}}
\end{figure}

MSPs are therefore difficult or impossible to detect at 430~MHz for the
high DMs usually found for the normal pulsar population near the
Galactic plane. Looking along the Galactic plane, the distance at
which a given DM is reached is much smaller than along lines-of-sight
at higher latitudes, this means that the volume within which MSPs can be
detected at 430 MHz is much smaller (both compared to normal pulsars,
and per square degree, henceforth indicated as $^{\Box}$) for the
low Galactic latitudes. This applies, to a lesser extent, to the
population of normal pulsars (see Fig.~\ref{fig:population}).

The high Galactic latitudes, with lines of sight containing
significantly less plasma, can be considered a better place
to look for MSPs from a point of view of their detectability.
The survey by Wolszczan (1990) has demonstrated, by finding a millisecond
pulsar orbited by the first known planets outside the solar system
(PSR~B1257+12) and a double neutron-star system (PSR~B1534+12) that
such studies are rewarding. This is confirmed by the large
number of MSPs found in the Parkes 70-cm all-sky survey
(Manchester et al. 1996, Lyne {et al.}  1998).

The scale height for the MSP distribution is of the order of 0.65
kpc (Cordes \& Chernoff 1997). A survey with good sensitivity can detect MSPs
at larger distances, and therefore more objects will be detectable
as we approach the plane of the Galaxy, because there will be
more pulsars along the line of sight. At some Galactic latitude $b$, the
number of MSPs will start to decrease as we approach the plane
because of the large amounts of plasma along the line of sight.

This intuitive result, depicted schematically in
Fig.~\ref{fig:population}, is also predicted by detailed simulations
carried out by one of us (JN, with Shri Kulkarni) and by more recent
simulations of the pulsar population of the Galaxy by Cordes and
Chernoff (1997)\nocite{cc97}, who estimate the optimal latitude
$b$ to be about $20^\circ$ at 430~MHz.

Motivated by our early simulations, we started a 430-MHz pilot
survey of a small region of the Galaxy with $7^\circ\,<\,|b|\,<\,20^\circ$
visible with the Arecibo radio telescope. Our aim was to test
the expectation that many MSPs remain to be discovered in these
intermediate Galactic latitudes. Some of the results of this
Intermediate Latitude Survey (ILS) are presented in this work. The
timing solutions for PSR~J1756+18, PSR~J2050+13 and PSR~J2016+1947,
and a detailed study of the latter pulsar will be presented elsewhere.

\section{Search observations and their sensitivity}\label{sec:obs}

\begin{figure}
\plotone{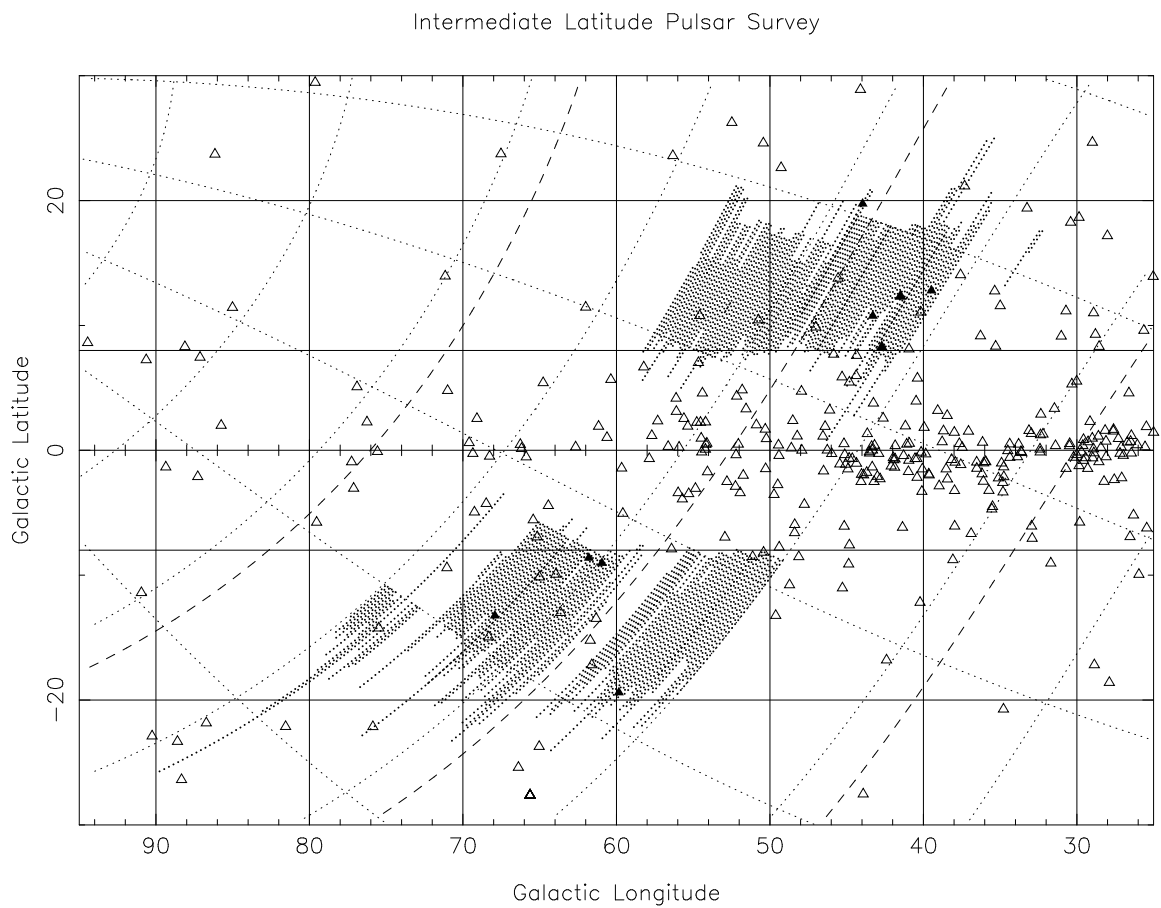}
\caption{Pointings made during the Intermediate Latitude Survey, in
Galactic coordinates. The unfilled triangles represent previously known
pulsars, the filled triangles represent the new pulsars. Each pointing
covers approximately the physical size of its representing dot at more
than 50\% of the sensitivity of the centre of the beam.
The dashed lines represent the southern declination limit, the zenith
declination and the northern declination limit of the Arecibo radio
telescope. The database for known pulsars is the ATNF pulsar
catalogue, which includes the pulsars found in the Parkes
multi-beam surveys
(http://www.atnf.csiro.au/research/pulsar/catalogue/). \label{fig:il1}}
\end{figure}

The ILS observations began in May 1989, and ended in
September 1991. They used the 430-MHz Carriage House line
feed. The data were acquired with the old Arecibo correlator, using
3-level quantization. The total bandwidth used was 10 MHz, the
number of lags being 128. The sampling time used was 506.625~$\mu$s,
each independent pointing containing $2^{17}$ such samples, for a
total integration of 66.4 seconds. The total number of pointings
processed was 6121, which
represents a total observing time of 113 hours and a total survey area
of about 130$^{\Box}$. The pointing positions are indicated in Figure
\ref{fig:il1} (Galactic coordinates).

The data were processed at Caltech and the Los Alamos National
Laboratories with the help of a Cray-YMP computer. The software used
for the data reduction was the Caltech pulsar package PSRPACK,
developed by Will Deich and one of us (JN). The multi-frequency data
were dedispersed at a set of 163 trial DM values between 0 and
200~cm$^{-3}$~pc. Each dedispersed time series was Fourier
transformed, and the strongest periodic components were recorded. In
order to increase sensitivity to signals with a small duty cycle,
harmonically related components (at $f$, 2$f$, ..., $nf$; with
$n\,=\,2$, 4, 8 and 16) were summed, and again the strongest peaks were
recorded. The power spectrum was also searched for harmonics up to
$n\,=\,4$ of frequency above Nyquist that might have been aliased back
into the power spectrum, in an attempt to improve our sensitivity to
millisecond pulsars.

The sensitivity of the ILS as a function of DM and period is
displayed in Fig.~\ref{fig:sens} for a pulsar with a square pulse
profile and a pulsewidth of 5\%. For this figure, we took into
account the observing system's equivalent flux density, integration
time and observing bandwidth, dispersive smearing and the effect of
the interstellar scattering as quantified by Cordes (2001).

The DM at which the sensitivity is degraded by half for a 5~ms pulsar
is about 150~cm$^{-3}$~pc. If we find a pulsar with this DM in the
direction of our survey for which the electron densities are
larger ($l\,=\,40^\circ$ and $b\,=\,7^\circ$); then such a pulsar is
at a distance of about 5 kpc, according to the Cordes and Lazio
model of the electron distribution in the Galaxy ({Cordes} \& {Lazio}
2002). The ILS can detect pulsars at these distances (e.g. PSR~J1819+1305).
For the direction where the electron densities are lower
($l\,=\,70^\circ$ and $b\,=\,-20^\circ$), no pulsar should be found
with such a large DM. The effect depends mainly on Galactic latitude
($b$). To summarize, for the higher Galactic latitudes the
detectability of MSPs at 430 MHz is limited only by the sensitivity of
the survey system, for the lower Galactic latitudes, it is also
limited by pulse smearing.

\section{Discoveries; pulsar timing}\label{sec:results}

\begin{table}
\begin{center}
\title{ Characteristics of the five newly found
pulsars with known phase-coherent timing solutions.}
\begin{tiny}
\begin{tabular}{l c c c c c c}
\hline
\hline
Pulsar   & 
J1814+1130 & 
J1819+1305 &
J1819+1305(a) &
J1828+1359 &
J2017+2043 & 
J2048+2255 \\
\hline 
Parameters: & & & & & & \\
Measured & & & & & & \\
\hline
Epoch (MJD) & 51500 & 51650 & 51650 & 51500 & 51500 & 51500 \\
Start (MJD) & 51207 & 51209 & - & 50901 & 50901 & 50901 \\
Finish (MJD) & 52645 & 52645 & - & 52637 & 52645 & 52645 \\
r.m.s. ($\mu$s) & 237 & 348 & - & 513 & 110 & 127 \\
N.TOAs & 101 & 95 & - & 110 & 82 & 103 \\
$\alpha$ (h:m:s) & 18:14:42.742(2) & 18:19:56.226(3) & 18:19:56.22(4) &
18:28:53.338(2) & 20:17:28.938(2) & 20:48:45.868(2) \\
$\delta$ ($^\circ$:':'') & 11:30:43.95(5) & 13:05:15.25(11) & 13:05:14.2(7) &
13:59:35.36(13) & 20:43:31.90(3) & 22:55:05:31(3) \\
$l$ ($^\circ$) & 39.20 & 41.23 & 41.2 & 43.02 & 61.38 & 67.45 \\
$b$ ($^\circ$) & 13.31 & 12.82 & 12.8 & 11.25 & $-$8.27 & $-$12.94 \\
Period (s) & 0.751261115038(3) & 1.060363543971(7) & 1.06036354400(6) &
0.741639520385(8) & 0.537143086032(2) &
0.2839009641977(10) \\
$\dot{P}$ ($10^{-15}$) & 1.66038(8) & 0.3592(2) & 0.373(9) & 0.7286(2) & 0.99555(5) & 0.01516(2) \\
DM (cm$^{-3}$\,pc) & 65 & 64.9 & 64.9 & 56 & 61.5 & 68.8 \\
$w_{50}$ (\%) & 1.3 & 6.3 & 5.9 & 1.7 & 0.9 & 2.2 \\
$S_{430}$ (mJy) & 0.72 & 6.2 & - & 1.2 & 1.5 & 1.8 \\
\hline
Derived & & & & & & \\
\hline
$\tau_c$ (Myr) & 7.2 & 45 & 45.0  & 16 & 8.5 & 300 \\
$B_0$ (Gauss) & 1.1$\,\times\,10^{12}$ & 6.3$\,\times\,10^{11}$ &
6.4$\,\times\,10^{11}$ & 7.4$\,\times\,10^{11}$ &
7.4$\,\times\,10^{11}$ & 6.6$\,\times\,10^{10}$ \\
$\dot{E}$ (erg s$^{-1}$) & 1.5$\,\times\,10^{32}$ &
1.2$\,\times\,10^{31}$ & 1.24$\,\times\,10^{31}$ &
7.1$\,\times\,10^{31}$ &
2.5$\,\times\,10^{32}$ & 2.6$\,\times\,10^{31}$ \\
Dist (kpc) & 2.7 & 5.1 & 5.1 & 3.0 & 3.4 & 4.2 \\
$L_{430}$ (mJy kpc$^2$) & 5 & 161 & - & 11 & 17 & 32 \\
\hline
\end{tabular}
\caption{\label{tab:parameters}
Timing and derived parameters for five
pulsars. The Right Ascension ($\alpha$) and the declination ($\delta$)
are indicated in J2000 coordinates.  (a) - as published in
({Edwards} {et al.}  2001) with the exception of the distance estimate.
Their value for the pulse duty cycle at half height
($w_{50}$) was measured at a frequency of 1400~MHz. The flux density
at 430~MHz ($S_{430}$) was calculated by averaging {\em all} the
detections of the pulsars made after obtaining their timing solution
(perfect pointing) and ignoring integrations with bad baselines.
The resulting pulse profile is then compared with the
the off-pulse r.m.s.; we assume the amplitude of the noise (in mJy)
to be the expectation of the the radiometer equation for the total
time added. The characteristic age, $\tau_c$, is
calculated using $\tau_c = P/(2 \dot{P})$, the surface magnetic flux
density $B_0$ is estimated using
$B_0\,=\,3.19 \times 10^{19} \sqrt{P \dot{P}}$ and
$\dot{E}\,=\,4\pi^2 I \dot{P} P^{-3}$, where $I$ is the moment of inertia
of the neutron star (assumed to be $10^{45}\,$g~cm$^2$). The distances
are estimated using using the latest electron model of the Galaxy
({Cordes} \& {Lazio} 2002). The uncertainties of the timing
parameters are twice the 1-$\sigma$ uncertainties obtained with
{\sc tempo}.}
\end{tiny}
\end{center}
\end{table}

\begin{figure}
\plotone{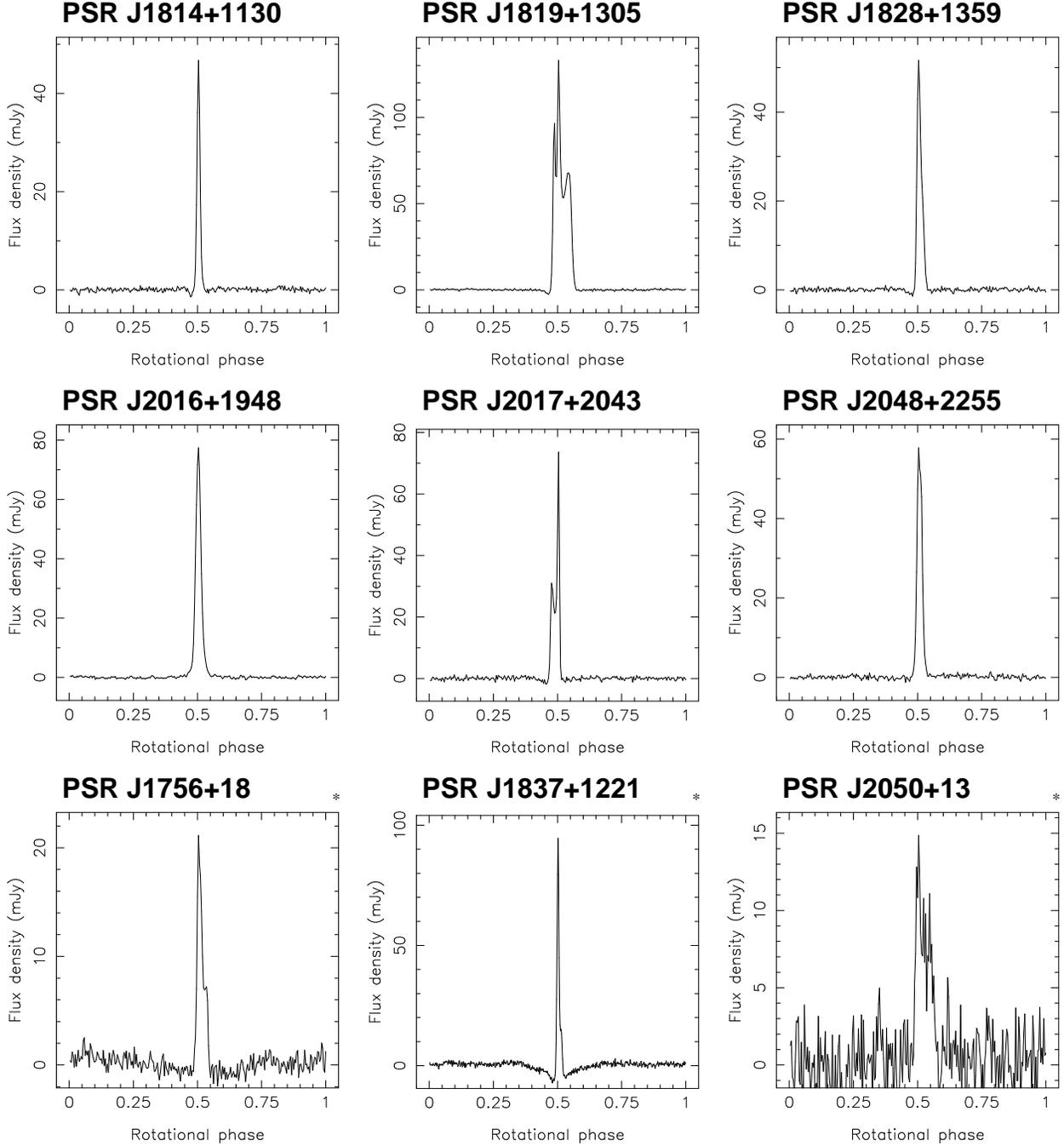}
\caption{Pulse profiles at 430 MHz for the nine pulsars discovered in
this survey, obtained by averaging the best detections. For the six
pulsars we have timed, these are also the templates used in deriving
the TOAs. The dips before and after the pulse profiles are caused by
the fact that the PSPM is a 4-bit machine that subtracts the radio
intensity signal from a running average, this significantly reduces
the amount of data to be acquired.\label{fig:profs}}
\end{figure}

The ILS detected a total of 13 pulsars, or about one per
10$^{\Box}$. Of these, four  were known before the start of our
survey: PSR~B1737+13, PSR~B1842+14, PSR~B2034+19 and PSR~B2053+21. The
remaining nine pulsars were previously unknown. Two of these were later
discovered and timed independently at Parkes: PSR~J1819+1305 and
PSR~J1837+1221 (Edwards et al. 2001). The pulse profiles for the 9 new
pulsars are presented in Figure \ref{fig:profs}.

Within the search area, there were 2 known pulsars that we did not
detect. One of them, PSR~J1838+16 (Xilouris et al. 2000), like seven
other pulsars found in the STScI/NAIC drift scan surveys and described
in that paper, has a flux density between 0.5 and 1 mJy. It is possible
that such a pulsar could have been missed because of interstellar
scintillation. The other pulsar, PSR~J2030+2228 (Rankin \& Benson 1981), has a
flux density at 400 MHz of about 5~mJy, i.e, we should expect for it a
S/N of about 140. It is not clear why this object was not detected, possible
causes are corruption of data with radio frequency interference (RFI)
or nulling.

Confirmation and timing of the new pulsars was done using 
the 430-MHz Carriage House line feed, just
as for the discovery observations. The re-observation attempts were made
in December 1997, and six new pulsars were confirmed then:
PSRs~J1814+1130, J1819+1305, J1828+1359, J2016+1948, J2017+2043 and
J2048+2255. None of the remaining seven candidates could be re-detected.
The back-end used to confirm these six pulsars and then to time them
 was the Penn State Pulsar Machine
(PSPM), a 128-channel filterbank. Each of the individual channels has
a bandwidth of 60 kHz, the sampling time is 80 $\mu$s and the data are
4-bit sampled. In timing mode, the PSPM-folded, multi-channel 
pulse profiles were dedispersed using SIGPROC (Lorimer 2001). The
topocentric times of arrival (TOAs) of the pulses at the telescope were
estimated using another routine from the same package.

The discovery at Parkes of PSR~J1837+1221, which coincides in
position, period and DM with one of our unconfirmed candidates,
led one of us (PCF) to re-observe the positions of
the previous candidates. So far, two more pulsars have been confirmed:
PSR~J1756+18 and PSR~J2050+13, again using the 430-MHz line feed and
the PSPM. It is unclear why these objects were missed in 1997, but
that might be because the telescope pointing had been
affected by the telescope upgrade works. One of the pulsars
(PSR~J2050+13) is also exceedingly faint, being sometimes undetectable
in 30-minute observations, it was found while scintillation was
amplifying its flux density.

\begin{table}
\begin{center}
\title{ Characteristics of three
pulsars without coherent timing solutions.}
\begin{tiny}
\begin{tabular}{l c c c}
\hline
\hline
Pulsar   & 
J1756+18 & 
J2016+1948 &
J2050+13 \\
\hline
Parameters: & & \\
Measured & & \\
\hline
$\alpha$ (h:m:s) & 17:56.0(3) & 20:16:56.7(5) & 20:50.0(3) \\
$\delta$ ($^\circ$:':'') & 18:19(5) & 19:48:03(7) & 13:01(5)\\
$l$ ($^\circ$) & 43.75 & 60.52 & 59.38 \\
$b$ ($^\circ$) & 20.23 & $-$8.68 & $-$19.11 \\
Period (s) & 0.744 & 0.0649403887(4) & 1.220 \\
DM (cm$^{-3}$\,pc) & 77 & 34 & 60 \\
$w_{50}$ (\%) & 2.1 & 2.1 & $\sim$ 2-3\\
$S_{430}$ (mJy) & $\sim$ 0.7 & 3.3 & $\sim$ 0.4 \\
$P_B$ (days) & - & 635.039(8) & - \\
$x$ (s) & - & 150.70(7) & - \\
$e$ & - & 0.00128(16) & - \\
$\omega$ ($^\circ$) & - & 90(5) & - \\
$T_{\rm asc}$ (MJD) & - & 51379.92(3) & - \\
\hline
Derived & & & \\
\hline
Dist (kpc) & 5.3 & 3.9 & 3.7 \\
$S_{430}$ (mJy kpc$^2$) & $\sim$19 & 35 & $\sim$5 \\
\hline
\end{tabular}
\caption{\label{tab:parameters1}Parameters for three pulsars without
phase-coherent timing solutions. See Table~\ref{tab:parameters} for
explanation of the parameters of the isolated pulsars. For
PSR~J2016+1948, $\alpha$ and $\delta$ are derived from gridding. This
pulsar's rotational period and its orbital parameters (orbital
period $P_B$, projection of the pulsar's orbital semi-major axis
along the line of sight, in
seconds, $x$, time of ascending node $T_{\rm asc}$, eccentricity $e$
and longitude of periastron $\omega$) and
their uncertainties are derived from a Monte-Carlo bootstrap
calculation. $T_{\rm asc}$ is preferred to the time of passage through
periastron because the orbit is nearly circular.}
\end{tiny}
\end{center}
\end{table}

Of the six pulsars we have timed since 1997, one, PSR~J2016+1948, is a
member of a 635-day pulsar-white dwarf binary system; it still has no
phase-connected timing solution. A consequence of this is that the
$\dot{P}$ has not yet been conclusively measured. For some of the more
recent measurements of this pulsar, we have also used the L-narrow
receiver at a central frequency of 1410 MHz, and the Wide-band Arecibo
Pulsar Processor correlator (WAPP) as a back-end with a total
bandwidth of 100 MHz. The signal-to-noise ratio obtained is similar to
that of the 430-MHz observations. The newly confirmed pulsars
PSR~J1756+18 and PSR~J2050+13 also lack timing solutions, and they are now
being timed at a frequency of 327 MHz. These three pulsars are listed
in Table \ref{tab:parameters1}. 

\begin{figure}
\plotone{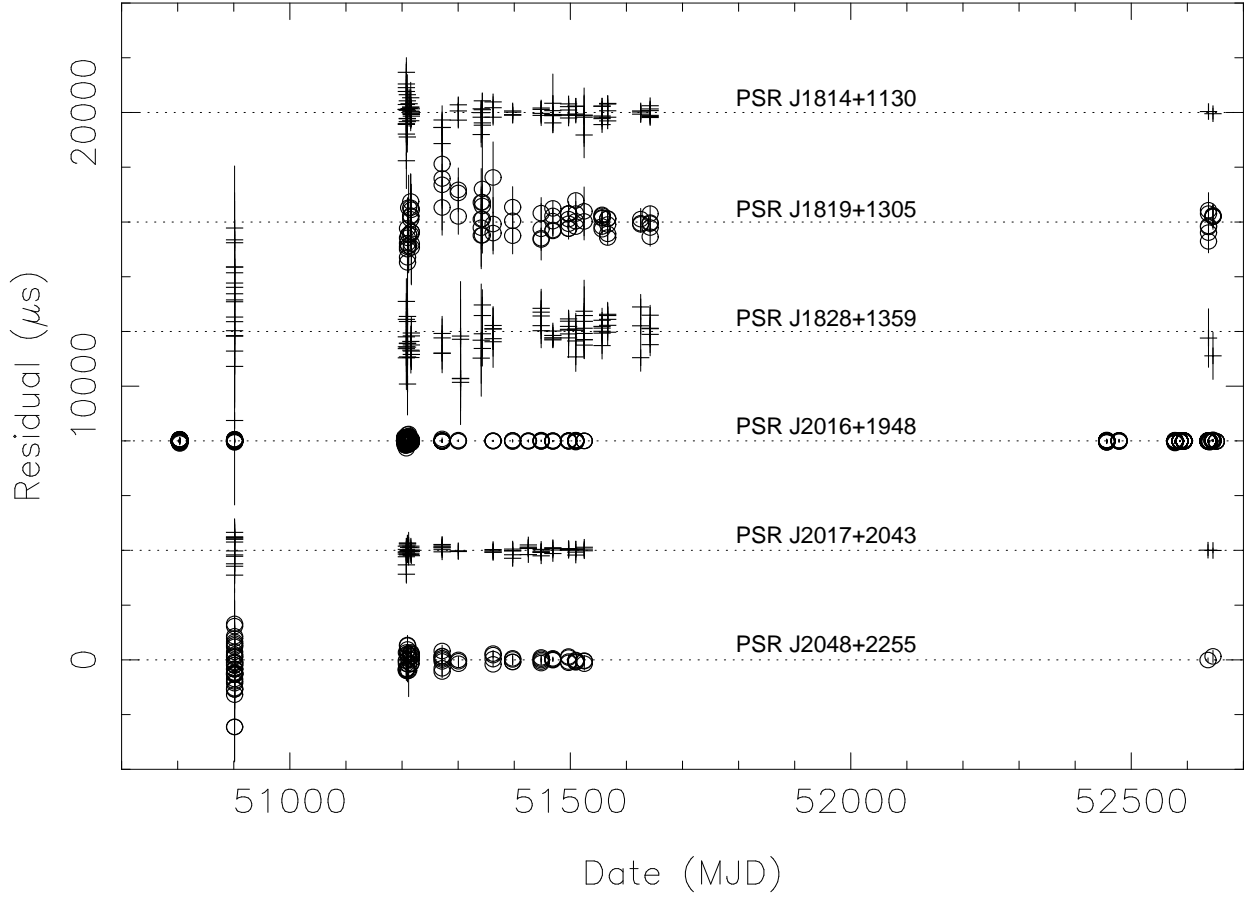}
\caption{Difference between the measured times of
arrival and the prediction of the timing models
for six of the newly discovered pulsars. These are displayed with
offsets of 20, 16, 12, 8, 4 and 0 ms for clarity.
There are no rotation ambiguities between the late 1990s data and the
2002/2003 data, the pre-fit phase error for the latter data was typically
less than 2\%. For PSR~J2016+1948 only the time coverage is indicated.
\label{fig:residuals}}
\end{figure}

The remaining 5 pulsars timed since 1997 have phase-coherent timing
solutions, which are presented in Table~\ref{tab:parameters}. These
were obtained and refined using the {\sc tempo} timing program
\footnote{http://pulsar.princeton.edu/tempo/}. Their timing parameters
are typical of the ``normal'' pulsar population; the characteristic
ages vary from 7 to 300 Myr. The residuals of the TOAs can be seen in
Figure \ref{fig:residuals}. For one of the two pulsars discovered
independently at Parkes, PSR~J1819+1305 (Edwards et al. 2001), the
timing parameters obtained at the two sites can be compared for the
same reference epoch (Table~\ref{tab:parameters}). The other pulsar
that was also found at Parkes (PSR~J1837+1221) has not been timed at
Arecibo.

\subsection{Understanding the survey results: Comparison with other surveys}

A direct comparison can be made between the ILS and the 1400-MHz
Swinburne Intermediate Latitude Survey (henceforth SILS),
which was made using the Parkes 64-m radio telescope's multi-beam system
({Edwards} {et al.}  2001). The SILS target area was defined by
$5^{\circ}\,<\,|b|\,<\,15^{\circ}$ and
$-100^{\circ}\,<\,l\,<\,50^{\circ}$, so there is some overlap with the
area covered by the ILS.

\begin{figure}
\plotone{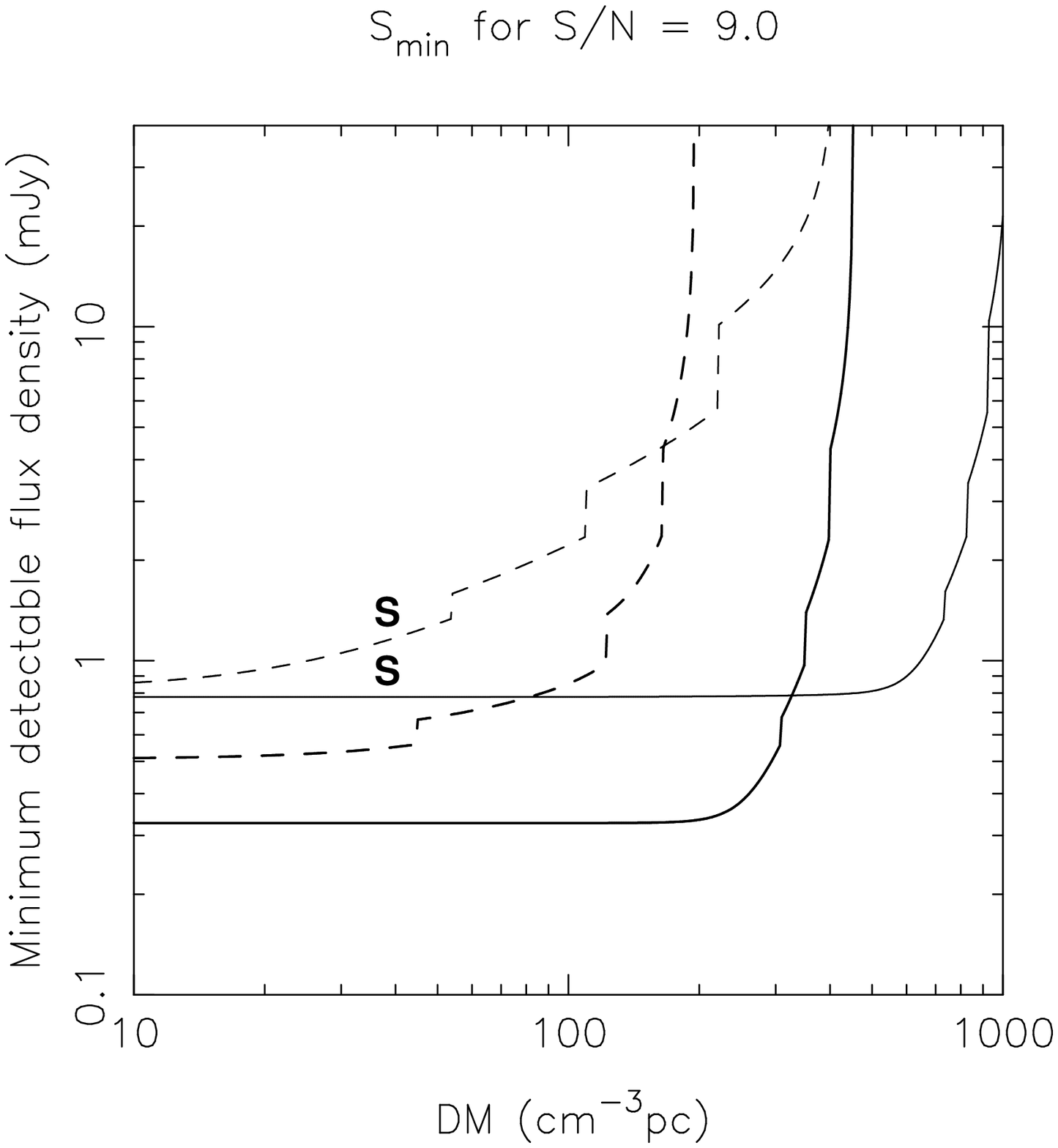}
\caption{Minimum detectable flux density of the
present survey (with S/N~=~9.0) for two pulsar periods, 5 ms (dashed
line) and 365 ms (solid line) as a function of DM, assuming a duty
cycle of 5\% and interstellar scattering as detailed in Cordes
(2001). The lighter curves, indicated with an ``S'', represent
the minimum flux density at 430~MHz of the weakest pulsars with the same
periods detectable by the Swinburne survey ({Edwards} {et al.}  2001),
assuming a relatively flat spectral index of
$-$1.2 (a more negative spectral index would cause an even greater
discrepancy in sensitivities between the two surveys).
The steps seen at high DMs are due to the harmonics of the
pulsar's fundamental frequency (due to each pulsar's peculiar pulse
profile, in this case a square wave) becoming undetectable, for
having a higher frequency than the time resolution of the experiment
at the given DM. \label{fig:sens}}
\end{figure}

In this overlap area, both surveys detected PSR~J1819+1305 and
PSR~J1837+1221. The SILS did not detect our two other discoveries in the
common search area, PSR~J1814+1130 and PSR~J1828+1359. Although we are
dealing with small number statistics, this result shows that the ILS
has achieved a greater sensitivity to normal pulsars in the
intermediate Galactic latitudes, as expected from Fig.~\ref{fig:sens}.

The SILS detected a total of 170 pulsars (69 of which were new
discoveries) in a search area 23 times larger than that of the ILS
($\sim 3000^{\Box}$). This represents one detection per $\sim
17^{\Box}$, which, as expected, is not as high as the detection
density of the ILS. However, the SILS covered
regions closer to the Galactic center and also at a slightly lower
latitude range ($5^\circ\,<\,|b|\,<\,15^\circ$), two factors that
increased that survey's detection rate.

The number of normal pulsars detected by the Swinburne survey
was not enhanced by the higher frequencies used. The pulsar with the
highest DM detected by the SILS, PSR~B1620$-$42, has a rotational
period of 0.365 seconds and a DM of 295~cm$^{-3}$~pc. Figure
\ref{fig:sens} shows that such a pulsar would not have been missed
by the ILS because of pulse smearing. Therefore, for the sensitivity
of the SILS, observing at 1400 MHz does not improve the
detectability of slow pulsars for the Galactic latitudes sampled.
This is acknowledged by
Edwards~et~al.~(2001), who find that the DM distribution of their
slow pulsar discoveries is similar to that of their re-detected
pulsars, which with almost no exception were found at lower
frequencies. We can therefore conclude that, as expected, our survey
is purely sensitivity-limited for this class of pulsars.

The situation changes for MSPs at similar DMs; where the SILS
is still capable of making detections, unlike any 430-MHz
survey (see Fig.~\ref{fig:sens}). The SILS has detected
twelve recycled objects, eight of them new (Edwards \& Bailes 2001), with DMs
between 26 and 117~cm$^{-3}$~pc. Fig.~\ref{fig:sens} shows that our
survey could in principle detect such objects. One consequence of this
is that the fraction of recycled pulsars detected by the
SILS (one in 14) is remarkably similar to ours (one in 13), even when
our survey finds twice as many normal and recycled pulsars per square
degree. Because of small number statistics, the recycled pulsar
fractions would still be consistent had we found one more or one less
recycled object. 

A more recent Arecibo 430-MHz survey of the Galactic plane (Nice,
Fruchter, \& Taylor 1995) has covered a region of the sky twice as
large (260$^\Box$, with $|b|\,<\,8^\circ$) with a sensitivity very similar
to that of the ILS. It detected 61 pulsars, of which 4 are
recycled. This represents more than twice the detection density of the
ILS survey, which is due to the larger concentration of pulsars along
the Galactic plane, yet the fraction of detections of recycled pulsars
(one in 15) is remarkably similar to that of the ILS and the SILS.
 It is, however, true that the fraction of recycled objects does
increase for the higher Galactic latitudes.

\subsection{Periodic nulling for PSR~J1819+1305}

PSR~J1819+1305 is by far the most luminous pulsar discovered in this
survey. Its pulse profile consists of three components;
the relative intensities of these are observed to vary systematically
from one 3-minute sub-integration to the next. Using single-pulse data
obtained in December 2002, we found that the main cause of this
variation is the strong intensity modulation of the first component
of the pulse profile.

\begin{figure}
\plotone{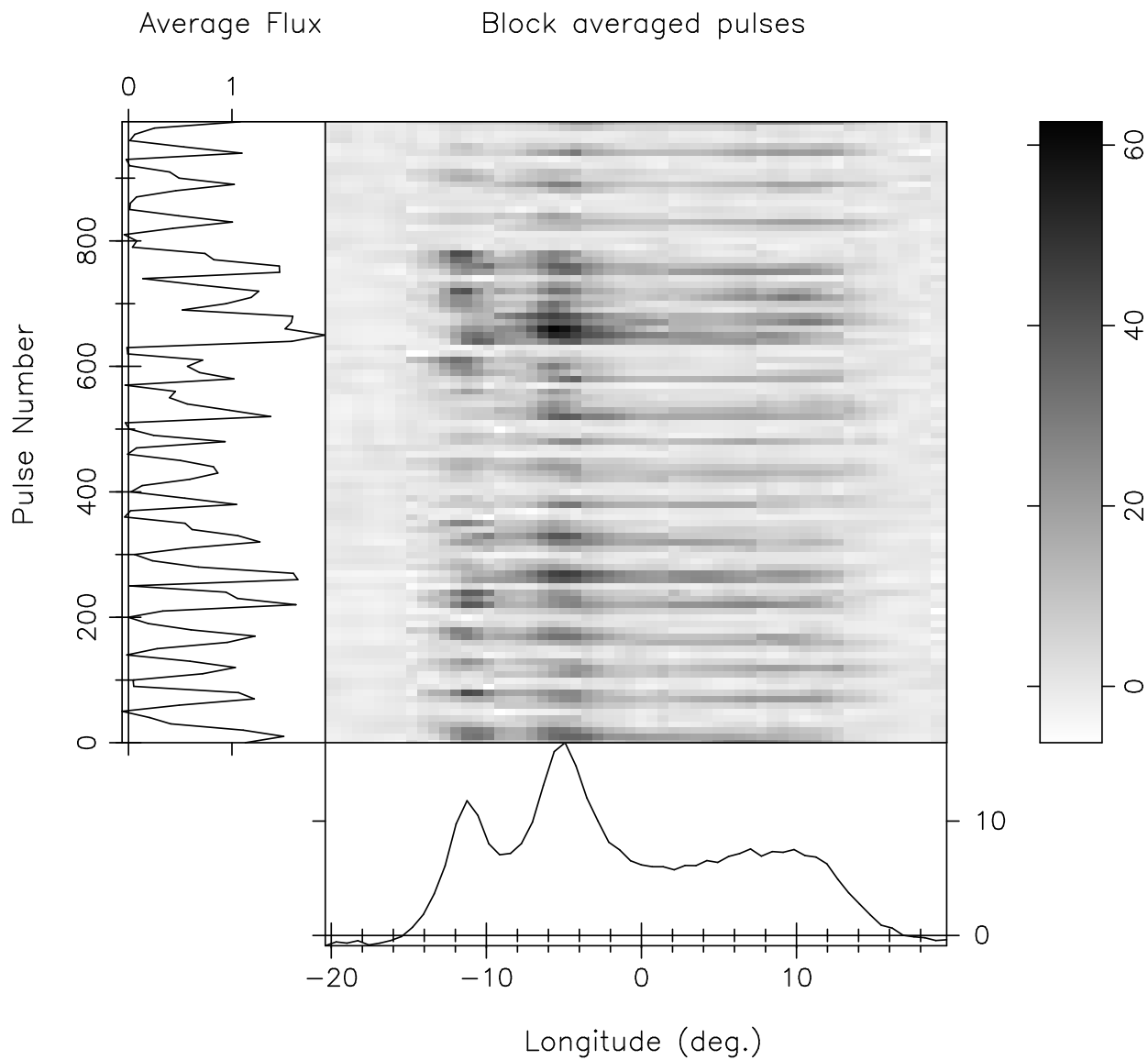}
\caption{A 17.2-minute observation of
PSR~J1819+1305 at 430 MHz. The data, obtained in December 29 2002,
were dedispersed at 64.9 cm$^{-3}$pc and folded according to the
ephemeris in Table~\ref{tab:parameters}. The grayscale plot represents
the pulsar's intensity as a function of rotational longitude
(horizontal axis, only a small portion of the whole cycle is visible)
and of pulse number (vertical axis). In the lower plot, we present the
integrated pulse profile. In the vertical plot on the left, we present
the pulse intensity as a function of pulse number. There the periodic
nature of the nulling becomes apparent. \label{fig:pulses}}
\end{figure}

These single-pulse data also showed that there is further intensity
modulation affecting the whole pulse profile, this can be seen in
Fig.~\ref{fig:pulses}.  Most of this is due
to  nulling, with emission absent for about
50\% of the time. This nulling has the peculiarity of having a very
marked periodicity at 53$\pm$3 rotations, which is both well defined,
long and with a large nulling fraction compared to other pulsars known
to exhibit periodic nulling (Rankin 1986).

The emission of this pulsar, by its combination of
peculiar characteristics, deserves a more careful study. In particular,
it will be interesting to
determine its polarization characteristics, which
might allow a good estimate of the angle between the magnetic and
rotation axis and of the latter relative to the line of sight. This
will be essential for a proper interpretation of the nulling, and
might lead to new insights about the emission mechanism of pulsars.

\subsection{The PSR~J2016+1948 binary system}\label{sec:2016+1948}

Perhaps the most important result of this work is the discovery of
the 65-ms pulsar PSR~J2016+1948. This is a member of a binary system,
together with a 0.29 M$_{\odot}$ white dwarf companion (assuming a
pulsar mass of 1.35~M$_{\odot}$ and an inclination of 90$^\circ$). The
orbital period is 635 days. PSR~J2016+1948 is the second most luminous
pulsar discovered in the ILS, with $L_{430}\,=\,35\,$mJy~kpc$^2$.

 For this binary system, we have not yet determined a phase-coherent
timing solution for the whole data set. The position in the sky was
determined with a set of 1400-MHz pointings at the source's nominal
position and half a beam width (one beam width is 3 arcmin at Arecibo)
north, south, east and west of the nominal position. A position can be
determined from the intensity of the pulsed signal of the detections,
with an uncertainty that is a small fraction of the beam size. This
procedure is known as ``gridding'' ({Morris} {et al.}  2002).

The remaining parameters for this pulsar were determined
from the observed barycentric periods. Technically, this was achieved
with TOA information: we have used {\sc TEMPO} to fit for the orbital
parameters and  rotational period with a different time offset for
each day's TOA set. Each of these sets gives an independent
estimate of the barycentric rotation period for its day. The orbital
model used was the ELL1 (Lange {et al.}  2001), which was specially
designed for low-eccentricity systems like PSR~J2016+1948, where no
precise estimates of the longitude of periastron (and therefore, of
the time of passage through periastron) can be made. We have used a
bootstrap Monte Carlo method ({Efron} \& {Tibshirani} 1993) to
estimate the 1-$\sigma$ orbital parameter uncertainties that appear in
Table \ref{tab:parameters1}.

 PSR~J2016+1948 has the third longest orbital period known for
this type of system; the longest are those of PSR~B0820+02, for which
$P_B\,=\,$1232 days (Arzoumanian 1995) and PSR~J0407+16 ($P_B\,=\,$669
days, $\dot{P}\,<\,10^{-18}$ and $e\,\sim\,0.001$,
Lorimer 2003, private communication).
 With a rotational period of 0.86~s, PSR~B0820+02
has not been significantly recycled by interaction with the
progenitor of the companion white dwarf. The much shorter rotational
periods of PSR~J2016+1948 (65 ms) and of PSR~J0407+16 (25 ms), and the
low upper limit for the period derivative of the latter pulsar
are suggestive of extensive recycling.

The determination of the orbital eccentricities for these systems is
important in its own right. The theories that describe the recycling
of neutron stars into MSPs (Alpar et al 1982, Zahn 1977, Phinney 1992)
predict the
order of magnitude of the orbital eccentricity as a function of the
orbital period. For $P_B\,=\,$600--700 days, the
eccentricity should be of the order of 10$^{-3}$ (Phinney 1992). The
values derived for the PSR~J2016+1948 and PSR~J0407+16
binary systems are in excellent agreement with that prediction.

\section{PSR~J2016+1948 as a gravitational laboratory}

Pulsars have been used in several different ways for testing the
fundamental properties of gravitation (Esposito-Far\`ese 1999, Bell
1999). One of the most fundamental and distinctive properties
of general relativity (GR) is the 
strong equivalence principle (SEP). Like the weak equivalence
principle (WEP), which led Einstein to elaborate GR, it requires
the universality of free fall: acceleration of any object in an
external gravitational field is independent of the size or chemical
composition of the object. However, SEP also requires the same
accelerations under external fields for objects that have
significantly different gravitational binding energies (Will 1993).

All theories of gravitation that describe gravity as a
distortion of space-time; the so-called metric theories of gravitation
(Will 1993), ``predict'' WEP by design. This is not the case for 
SEP, which is a feature peculiar to GR. If the assumption of SEP is
wrong, as postulated in many alternative theories of gravitation, i.e., if
\begin{equation}
\frac{m_I}{m_G} - 1 \equiv \Delta = \eta \times \frac{U_G}{m_G\,c^2} \neq 0,
\end{equation}
(where $m_I$ and $m_G$ are the inertial and gravitational masses of an
object, $\eta$ is the Nordtvedt parameter which measures deviations
from SEP, $U_G$ is the object's self-gravitational energy and $c$ is
the velocity of light) then the accelerations in
the same external field of two objects will not be exactly equal
because of the different gravitational binding energies.

This difference in acceleration causes a ``polarization'' of the
binary, which is an increase in the eccentricity of the system along
the direction of the external field known as the ``Nordtvedt effect''
(Nordtvedt 1968a).

Such an effect has not been found in the weak fields probed by
solar system experiments, in particular the Lunar Laser Ranging
(Nordtvedt 1968b). This experiment determined that, despite the
differences in gravitational self-energy between the Earth and the
Moon, these two objects fall at the same rate (to within one part in
10$^{13}$) in the gravitational field of the Sun ({Williams}, {Newhall}, \& {Dickey} 1996).
This implies that $\eta \, = \,- 0.0007 \,\pm \, 0.0010$, 
which is entirely consistent with GR. However, Damour and
Esposito-Far\'{e}se (1992) have shown that, generally,

\begin{equation}
\eta\,=\,\eta_W\,+\,\eta_S (c_1\,+\,c_2\,+\, ...)\,+\,...
\end{equation}

where $\eta_W$ and $\eta_S$ are the weak and strong field components
of the Nordtvedt parameter $\eta$ and the $c_i$ are the compactness of
the bodies involved:

\begin{equation}
c_i\, = \,\frac{U_{G,i}}{M_i c^2}.
\end{equation}

Is there a strong-field component of $\eta$? If so, GR is not the
right description of gravitation. Such a test cannot be conducted in
the solar system, where no strong fields are to be found.

A pulsar- white dwarf system, with the Galaxy generating the
external field, provides the
ideal laboratory to make such a measurement. Pulsars have very large
gravitational binding energies of about $-$15\% of the total mass
(i.e., $c_1 \,\sim\,-0.15$, the exact number depends on the equation
of state for cold matter at high densities and the mass of the
pulsar). The white dwarf companion has, comparatively, a negligible
gravitational binding
energy, about 10$^{4}$ times smaller than that of the neutron star. If
there is any strong-field component of the acceleration, it should be
felt just by the pulsar, hence the difference in acceleration compared
to the white dwarf and the associated Nordtvedt effect.

The figure-of-merit of a binary system for a Nordtvedt test is $P_B^2/e$
(Arzoumanian 1995). However, Wex (1997) has shown that for an
unambiguous interpretation of the low eccentricity of a binary system,
the system's age has to be significantly larger than one Galactic
orbit.

Among all the known binaries with previously published timing
solutions, the $P_B^2/e$ number is highest for
the PSR~B1800$-$27 system. The eccentricity of this
system cannot be interpreted unambiguously because of the pulsar's age
($\tau_c\,=\,$300 Myr) is of the order of a single Galactic orbit. 
Of all such binaries that pass the large
$\tau_c$ criterion, PSR~J1713+0747, with $\tau_c\,=\,8.5\,$Gyr
(Camilo, Foster, \& Wolszczan 1994), is the system for which $P_B^2/e$
is largest: 6.14$\,\times\,10^{7}$~days$^2$. For PSR~J2016+1948 and
PSR~J0407+16, the $P_B^2/e$ is $\sim 3\,\times\,10^{8}$~days$^2$ and
$\sim 5\,\times\,10^{8}$~days$^2$, or factors of $\sim$4-5 and 7-8 
times larger than for PSR~J1713+0747.

Using all the binary systems that pass the large age criterion, a
value of $|\Delta|\,<\,0.004$ was obtained (Wex 1997). Assuming
again that the pulsar's compactness is 0.15, this implies
$\eta\,<\,0.027$. The inclusion of PSR~J2016+1948 and PSR~J0407+16
in that ensemble of binary systems will significantly reduce the upper
limits on $|\Delta|$ and $\eta$.
An interesting aspect of these limits is that they will diminish
with the mere addition of binary systems with high
$P_B^2/e$ to the ensemble being used (Wex 1997).

The upper limit of $|\Delta|$ has been used, together with the
measurements of orbital decay for the
PSR~B1913+16 binary system, to impose fundamental
constraints to any alternative theories of gravitation, in particular
the tensor bi-scalar theories (Damour \& Esposito-Far\`ese 1992,
Esposito-Far\`ese 1999). These constraints
will become significantly more stringent with the inclusion of systems
like PSR~J2016+1948 and PSR~J0407+16.
However, we must keep in mind that the determination
of the equivalence of inertial and gravitational mass, particularly
when very large self-gravitational energies are
involved, is an important measurement in itself, with a significance
wider than the tests it introduces to any particular gravitational
theory. It will forever remain as a fundamental constraint to our
understanding of gravitation.

\section{Conclusions}\label{sec:conclusions}

We have found nine new pulsars in a small Arecibo 430-MHz survey of the
intermediate Galactic latitudes. We have timed six of these,
with five now having phase-coherent timing solutions. These are
old, normal pulsars; a
population similar to that of the earlier Hulse-Taylor survey. As
expected, there are no young pulsars among the sample that has been timed.
 We compare this survey with the SILS ({Edwards} {et al.} 2001) and
conclude that we have reached greater sensitivity to this population
of normal pulsars. 

Two of the discoveries, PSR~J1819+1305 and PSR~J1837+1221, were
independently found at Parkes by the SILS and then timed.
We found that the former pulsar has strong variation
of its integrated pulse profiles. This is partly due to nulling;
the nulls exhibit a very strong periodicity at 53$\pm$3 rotations.

Because of small number statistics, it is impossible to obtain
any firm conclusions, based on our survey alone, as to the number of
recycled pulsars to be discovered at the intermediate Galactic
latitudes at 430~MHz. We found a single pulsar that is
likely to be recycled, PSR~J2016+1948, out of a total of 13
detections, a proportion similar to what was obtained by 430-MHz
surveys of the Galactic plane. However, a much larger 430-MHz ILS
survey would certainly not find a larger fraction of recycled
pulsars than the SILS at 1400 MHz (which is also one in
14). Therefore, our initial expectation, and also that of Cordes and
Chernoff (1997), that for the latitudes near $|b| = 20^\circ$
there are many recycled pulsars to be found (this being the main motivation
of the present survey) cannot be confirmed by the ILS and SILS. This
might imply that the scale height for recycled pulsars is larger than
the value that was assumed in the previous calculations, 0.65 kpc.

PSR~J2016+1947 has a rotational period of 64.9 ms and it forms
a binary system with an $\sim$0.3~M$_{\odot}$ white dwarf
companion. This system  has an orbital period of 635 days, with an
orbital eccentricity of about
0.0012. The more precise determination of the orbital parameters of
this binary will, together with the timing of another binary pulsar,
PSR~J0407+16, lead to much tighter constraints on any violation of the
Strong Equivalence Principle.

\acknowledgements

The Arecibo Observatory, a facility of the National Astronomy and
Ionosphere Center, is operated by Cornell University under a
cooperative agreement with the National Science Foundation. The Los
Alamos National Laboratory (LANL) is operated by the University of
California for the National Nuclear Security Administration. We wish
to thank Jon Middleditch (LANL) for his hospitality and for making
some of LANL's computer resources available to us; Will Deich
(formerly Caltech, now Lick Observatory) for allowing us to use some
of his software; Alex Wolszczan for making the PSPM, the instrument
used to time the pulsars mentioned in this paper, freely available for
use at the Arecibo Observatory; Duncan Lorimer, for comments and ideas
that greatly improved the quality of this work and for making his
pulsar processing software publicly available, and Chris Salter and
Avinash Deshpande for comments that improved the quality of the
manuscript. Avinash Deshpande also helped with the preliminary
analysis of the PSR~J1819+1305 data.

\pagebreak

\end{document}